\documentclass[aps,prb,twocolumn,superscriptaddress]{revtex4}%
\usepackage{epsfig,pslatex,latexsym,times,amssymb,amsmath,graphicx}
\usepackage{bm, float, lipsum}
\usepackage{amsmath}
\usepackage{amsfonts}
\usepackage{amssymb}
\usepackage{mathrsfs}
\usepackage{color}
\usepackage{graphicx}%

\usepackage[normalem]{ulem}

\providecommand{\U}[1]{\protect\rule{.1in}{.1in}}
\begin{document}
\author{N. J. Harmon}
\email{harmon.nicholas@gmail.com} 
\affiliation{Department of Physics and Astronomy and Optical Science and Technology Center, University of Iowa, Iowa City, Iowa
52242, USA}
\affiliation{Department of Physics and Engineering Science, Coastal Carolina University, Conway, South Carolina
29528, USA}
\author{M. E. Flatt\'e}
\email{michael_flatte@mailaps.org} 
\affiliation{Department of Physics and Astronomy and Optical Science and Technology Center, University of Iowa, Iowa City, Iowa
52242, USA}
\affiliation{Department of Applied Physics, Eindhoven University of Technology, Eindhoven 5612 AZ, The Netherlands}
\date{\today}
\title{Driving a pure spin current from nuclear-polarization gradients}
\begin{abstract}
A pure spin current is predicted to occur when an external magnetic field and a linearly inhomogeneous  spin-only field are appropriately aligned. Under these conditions (such as originate from nuclear contact hyperfine fields that do not affect orbital motion) a linear, spin-dependent dispersion for free electrons emerges from the Landau Hamiltonian. The result is that spins of opposite orientation flow in opposite directions giving rise to a pure spin current.  A classical model of the spin and charge dynamics reveals intuitive aspects of the full quantum mechanical solution. We propose optical orientation or electrical polarization experiments to demonstrate this outcome. 
\end{abstract}
\maketitle	

\section{Introduction} 
The coupling of spin and orbital currents is integral to spintronics\cite{Awschalom2002,Awschalom2007}. The (inverse) spin Hall effect is a hallmark example where (spin) charge current is converted to (charge) spin current.\cite{Dyakonov1971a, Dyakonov1971, Kato2004b, Wunderlich2005} Other effects include: spin galvanic or ``Edelstein'' effects (and their reciprocal) which convert charge current into spin polarization.\cite{Edelstein1990, Ganichev2002, Sanchez2013} Each of these rely on the intrinsic coupling of spin and charge via the spin-orbit effect. Despite this there are a few examples of spin-charge current coupling not through spin-orbit effects, such as the spin Gunn effect\cite{Qi2006a,Qi2006b} or spin bottleneck effects in localized\cite{Petta2005} or extended\cite{Bobbert2007,Harmon2012a} materials, which rely on the Pauli exclusion principle and dynamical spin correlations. 
\begin{figure}[hh]
 \begin{centering}
        \includegraphics[scale = 0.45,trim = 270 5 200 110, angle = -0,clip]{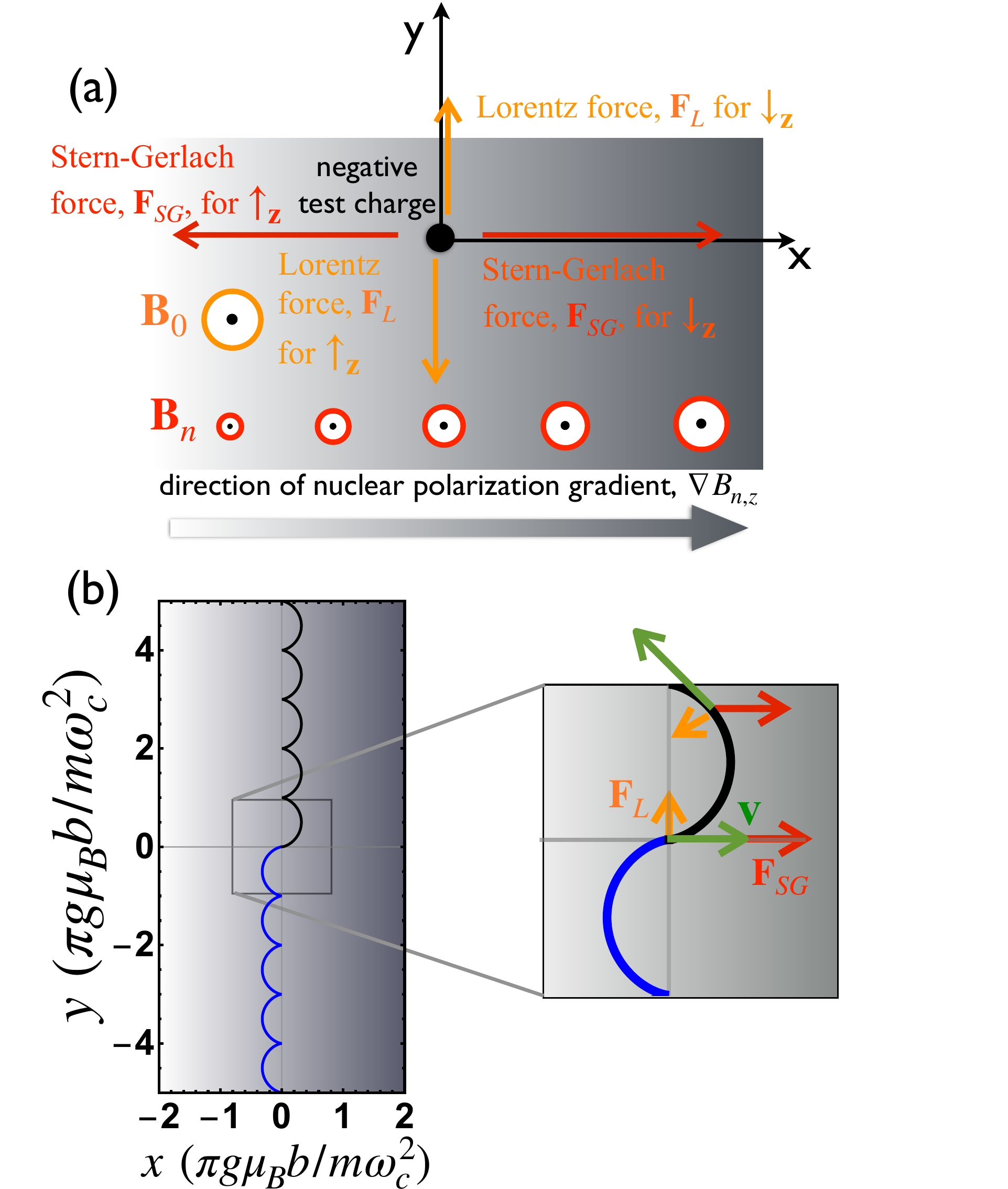}
        \caption[]
{Diagram of transverse geometry in $n$-GaAs where gradient force is perpendicular to applied field. 
Trajectories of Eqs. (\ref{eq:traj1}, \ref{eq:traj2}) are shown. Regardless of initial conditions, spins travel in opposite directions at the same speed. Blue trajectory is up spin and black trajectory is down spin. Inset: orange arrows represent Lorentz force, red arrows represent Stern-Gerlach force, and green arrows represent charge velocity.}\label{fig:geometry} 
        \end{centering}
\end{figure}

In this article an alternate method of spin-charge coupling is described that relies on electron-nuclear spin coupling and does not require   spin-orbit coupling,  the Pauli exclusion principle, or electronic spin-spin correlations.

The origin of the effect is dynamic nuclear polarization: large nuclear spin polarizations that can exert a sizable nuclear field on the electronic system\cite{Overhauser1953, Meier1984}. These nuclear spin polarizations are generated by electron non-equilibrium spin transfers to moment-carrying nuclei through the hyperfine interaction which accumulate due to the slow spin relaxation time of nuclei. 
The resultant nuclear field, although it is a magnetic field, is highly concentrated near the nuclei (Fermi contact potential).  Due to its localized character and the lack of an extended vector potential acting on the orbital motion, this nuclear field acts only on spin and not orbital motion, and is sometimes referred to as an \emph{effective} field. Here we designate it as a ``Zeeman-only field".

This absence of coupling between nuclear spin and electronic orbital momentum entails the field from polarized nuclei does not contribute to the ordinary Hall effect but may support a larger anomalous Hall effect due to the increase in spin splitting \cite{Bednarski1998, Vagner2004}.
How a Zeeman-only field allows spin components to be spatially separated is the subject of this article.

Here we show that nuclei with patterned spin polarization, due to their lack of orbital coupling and spatial inhomogeneity, can evince remarkable spin-dependent charge dynamics leading to pure spin currents or charge currents.
Recent work has demonstrated the importance of inhomogeneous nuclear fields on coupled electron-nuclear spin dynamics \cite{Harmon2015, Ou2016} but have not explored the resulting spin-dependent motion.
The spin-motive force may emerge from either the Stern-Gerlach force of from the combination of Stern-Gerlach and Lorentz forces.
The effect of the net force is to separate up and down spins along a direction longitudinal or transverse to the effective field gradient (Fig. \ref{fig:geometry}(b) shows spins separating transverse to the gradient). In a longitudinal field configuration, a linear contribution to the dispersion relation appears for which we calculate a spin current using a Landau-like Hamiltonian. We also examine the longitudinal and transverse geometries within a semiclassical Drude-like model.
Finally, we propose experiments to see this effect by inducing dynamic nuclear polarization gradients via patterned optical or electrical orientation.


\section{Nuclear field} 
We treat an electron ensemble in two or three dimensions with charge $q = -e$, effective electron mass $m$, and effective Land\'e $g$-factor $g^*$.
We assume that the external field, $\bm{B}_{0}$, is homogeneous and any inhomogeneities lie in a Zeeman-only field, $\bm{B}_{\mathcal{Z}}(\bm{r})$.
We choose a Zeeman field based on the magnetic interaction between electrons and nuclei:
$\bm{B}_{\mathcal{Z}}(\bm{r}) = \bm{B}_n(\bm{r}) $ where
\begin{equation}\label{eq:nucfield}
\bm{B}_n(\bm{r})  = b_{n}  \langle \bm{I}(\bm{r}) \rangle  = b_{n}  \langle I (\bm{r})\rangle \hat{B}_{0},
\end{equation}
with 
\begin{equation}\label{eq:zeemanfield}
\langle I (\bm{r})\rangle = \frac{ B_{0}^2}{B_{0}^2 + \xi B_{\ell}^2}\bm{P}(\bm{r}) \cdot \hat{B}_{0}.
\end{equation}
Here $b_{n}$ is the Overhauser coefficient and $\bm{P}(\bm{r}) = \bm{P}_0 + \bm{\beta} f(\bm{r}) = P_0 \hat{B}_{0}+ \beta f(\bm{r})\hat{B}_{0}$  is the out-of-equilibrium electron spin polarization (we assume that thermal electron spin polarization is negligible) that is responsible for the dynamic nuclear polarization.
The quantity $\beta$ controls the magnitude of the non-uniformity whereas $f(\bm{r})$ is the function specifying the spatial structure of the Zeeman-only field; $\sqrt{\xi}B_{\ell}$ is the strength of the random local field \cite{Paget1977}.
A coordinate system is chosen to maintain the external field in the $z$-direction. 
$\bm{B}_n(\bm{r})$ lies collinear with $\bm{B}_{0}$ (we ignore Knight fields) but its gradient may not; we examine two different \emph{linear} functions for $f(\bm{r})$: longitudinal [$f(\bm{r}) = z$] and transverse [$f(\bm{r}) = x$]. These linear functional forms describe slowly varying exponential functions ($-e^{-r_i} \sim r_i-1$) of their respective Cartesian coordinate, $i \in \{x,y,z \}$, so that we can posit the existence of an constant spin-dependent force as shown later in this article.
Combining these assumptions, the nuclear field is
\begin{equation}
\bm{B}_n(\bm{r})  = c \hat{B}_0 + b r_i \hat{B}_0
\end{equation}
where 
\begin{equation}
b = b_n \frac{ B_{0}^2}{B_{0}^2 + \xi B_{\ell}^2 } \beta, \qquad c =b_n \frac{ B_{0}^2}{B_{0}^2 + \xi B_{\ell}^2 }P_0.
\end{equation}

\section{Quantum mechanical formulation} 
The starting point is the Landau level description of free electrons in a magnetic field.
The Hamiltonian for Landau levels is
\begin{equation}\label{eq:LandauLevel}
\mathscr{H} = \frac{p_x^2}{2 m} + \frac{1}{2}m \omega_c^2(x - x_0)^2
\end{equation}
where the Landau gauge, $\bm{A} = (0, B_0 x, 0)$ is assumed and $\omega_c = e B_0/m$. The harmonic potential is centered at $x_0 = -\hbar k_y/e B_0$.
The lack of propagation in the bulk is shown by a quick computation of the velocity $v_y = \hbar k_y/m - q A_y(x_0)/m = \hbar k_y/m - q B_0 x_0 /m = 0$.
The wave functions are
\begin{equation}
\Psi(\bm{r},x_0)  =  \frac{e^{i (k_y y+k_z z)}}{\sqrt{2^n n!}}\left(\frac{m \omega_c}{\pi \hbar}\right)^{1/4} e^{-\frac{m \omega_c (x-x_0)^2}{2 \hbar}}H_n\left(\sqrt{\frac{m \omega_c}{\hbar}}(x-x_0) \right)
\end{equation}
where $H_n$ are Hermite polynomials of degree $n$.
To the best of our knowledge, prior work has not treated effective or Zeeman-only magnetic field gradients within the Landau Hamiltonian.
Inhomogeneous \emph{real} magnetic fields, that operate on spin and orbital degrees of freedom, lead to more challenging Hamiltonians that exclude analytic solutions.
For instance, a linear magnetic field gradient produces an anharmonic oscillator potential ($\sim (x^2 - x_0^2)^2$) for which there is no analytic solution \cite{Muller1992}.
By assuming a constant real magnetic field in $z$ and a linearly inhomogeneous Zeeman-only field (directed in $z$ but changes along $x$ --- we call this the transverse geometry), difficulties are avoided since the Landau level Hamiltonian is unchanged from Eq.~(\ref{eq:LandauLevel}) except for the addition of a Zeeman term that contains the field inhomogeneity:
\begin{equation}\label{}
\mathscr{H} = \frac{p_x^2+p_y^2+p_z^2}{2 m} + \frac{1}{2}m \omega_c^2(x - x_0)^2 + \frac{g e}{2 m} (B_0 + c + b x)S_z.
\end{equation}
Since the equation depends on neither $y$ nor $z$, we express those dependencies of the wave function as planes waves which leaves us, for each spin orientation $\sigma = \pm 1$, with, after ``completing the square" and dropping terms of order $b^2$, with
\begin{widetext}
\begin{equation}\label{}
\frac{-\hbar^2}{2 m}\frac{\partial^2}{\partial x^2} \phi(x) + \frac{1}{2}m \omega_c^2\big[x - x_0'\big]^2 \phi(x) = \Big[\varepsilon - \frac{ \hbar^2k_z^2 }{2m} - \frac{g \mu_b}{2}(B_0 + c)\sigma - \frac{1}{2} g \mu_B b x_0 \sigma \Big] \phi(x),
\end{equation}
\end{widetext}
where $x_0' = x_0 - \frac{g \mu_B b}{2 m\omega_c^2}\sigma$  is the new center of the harmonic potential and $\phi(x)$ is the $x$ part of the separable wave function.
The eigenvalues of this modified Landau problem are $E = \hbar \omega_c (n + \frac{1}{2})$ which gives a total energy of
\begin{equation}
\varepsilon = \hbar \omega_c (n + \frac{1}{2}) + \frac{\hbar^2 k_z^2 }{2m} + \frac{g \mu_b}{2}(B_0 + c)\sigma - \frac{1}{2} g \mu_B b \frac{\hbar k_y}{e B_0} \sigma,
\end{equation}
which possesses a linear-in-$k_y$ dispersion. 
The wave function differs only slightly from the Landau level case: $\Psi(\bm{r}, x_0')$.

The group velocity is defined as $\bm{v}_g = \partial \varepsilon/\hbar\partial \bm{k}$.
$v_{g,x}$ is trivially zero and $v_{g,z}$ is $\hbar k_z/m$ but on average also zero since $\int_{-\infty}^{\infty} k_z dk_z = 0$. There is no type of current in $x$ or $z$.
However the linear term remains for the $y$ group velocity: $v_{g,y} = \partial \varepsilon/\hbar\partial k_y = -\frac{g \mu_B b}{2 e B_0} \sigma =  -r_{SG} \omega_c \sigma$ where $r_{SG} = g  \mu_B b /2 m \omega_c^2$ and $\omega_c = e B_0/m$.
As expected, different spin orientations move in opposite directions.
Summing over the two spins yields zero charge current. 
If only a single spin orientation were present, then a charge current would accompany the spin polarized current. 

A harmonic confining potential can be added to mimic edges ($V_{confine} = m \omega_0^2 x^2/2$), and remarkably the problem can still be solved exactly;
$x_0$ becomes 
\begin{equation}
x_0 \rightarrow \frac{\omega_c^2}{\omega_c^2+\omega_0^2} x_0'= \frac{\omega_c^2}{\omega_c^2+\omega_0^2}\big( x_0 - \frac{g \mu_ b b}{2 m \omega_c^2}\sigma\big)
\end{equation}
and the eigenvalues are
\begin{widetext}
\begin{equation}
\varepsilon = (n + \frac{1}{2}) \hbar (\omega_c^2+\omega_0^2)^{1/2}  - \frac{1}{2} \frac{g \mu_B b \hbar k_y \omega_c}{m (\omega_c^2+\omega_0^2)} \sigma + \frac{\hbar^2 k_y^2}{2m}\frac{\omega_0^2}{\omega_c^2 + \omega_0^2}  + \frac{\hbar^2k_z^2 }{2m} + \frac{g \mu_b}{2}(B_0 + c)\sigma 
\end{equation}
\end{widetext}
where the main difference between the unconfined example is the presence of a kinetic energy with a modified effective mass (3rd term).
The dispersion relation contains spin-dependent linear and spin-independent quadratic elements.
The eigenstates are not significantly altered beyond a redefinition of $x_0$ and a new normalization factor \cite{Berggren1988}.
Confinement does not change the results in any significant way --- a pure spin current is still generated traveling in the $\mp y$ direction:
\begin{equation}
v_{g,y} = -\frac{1}{2}\frac{g \mu_B b \omega_c}{m (\omega_c^2 + \omega_0^2)}\sigma,
\end{equation}
where the effect of the harmonic potential is to reduce the velocity.

The presence of the soft potential allows us to avoid the unphysical fact that the velocity diverges as $\omega_c \rightarrow 0$ in the unconfined model. With the soft potential in place, the transverse velocity also vanishes if the applied field vanishes in accordance with expectations.

\section{Semi-Classical formulation} 
The spin separation is naturally seen within a simple classical framework that includes discrete spins.
Only real fields exert a Lorentz force, $\bm{F}_L = -e \bm{v} \times \bm{B}_{0}$, while the gradient of either field ($\bm{B}_{0}$ or $\bm{B}_{n}$) may exert a Stern-Gerlach force; since $\bm{B}_{0}$ is uniform, the spin-dependent forces are
 \begin{equation}\label{eq:iR}
\bm{F}_{SG} = -\nabla (-\bm{\mu} \cdot \bm{B}) = -g\frac{\mu_B }{\hbar}  \nabla (\bm{S} \cdot \bm{B}) = - \frac{g}{2} \mu_B \sigma  b \hat{z}, ~ \text{(longitudinal)}
 \end{equation}
 \begin{equation}\label{eq:iR2}
\bm{F}_{SG} = -\nabla (-\bm{\mu} \cdot \bm{B}) = -g\frac{\mu_B }{\hbar}  \nabla (\bm{S} \cdot \bm{B}) = - \frac{g}{2} \mu_B \sigma  b \hat{x}, ~ \text{(transverse)}
 \end{equation}
for gradients either longitudinal or transverse to $\bm{B}_{0}$.
Our choice of Zeeman-only field along $\hat{z}$ allows the spin dynamics to be trivial when enforcing semiclassical spins to be in one of two states $\bm{S} = \frac{\hbar}{2}(0, 0, \sigma)$ where $\sigma = \pm 1$.
The charge and spin dynamics are determined by solving the equations of motion:
\begin{equation}
\bm{F}_L + \bm{F}_{SG} = m\frac{d \bm{v}}{dt} = -e B_0 \bm{v} \times \hat{z} - \frac{g}{2}\mu_B \sigma  b  \hat{z}, ~ \text{(longitudinal)}
 \end{equation}
 \begin{equation}
\bm{F}_L + \bm{F}_{SG} = m\frac{d \bm{v}}{dt} = -e B_0 \bm{v} \times \hat{z} - \frac{g}{2}\mu_B \sigma  b   \hat{x}. ~ \text{(transverse)}
 \end{equation}
In either case, the constant force acts just like a spin-dependent effective constant electric field.
In the longitudinal geometry, consisting of a constant force, the charge carrier accelerates indefinitely.
By considering damping (to be done in next section), this unphysical behavior is avoided.
The system of equations for the transverse model can be solved exactly for any initial starting place and electron velocity in a way that mirrors the classical Hall effect calculation except now with a spin-dependent electric field.
For simplicity we express the solution for an electron starting at the origin with no initial velocity, $v_0 = 0$,
\begin{equation}\label{eq:traj1}
\bm{r}(t) = \bigg(- \sigma r_{SG} (1 - \cos\omega_c t),
 \sigma r_{SG}  ( -\omega_c t +  \sin \omega_c t), 0\bigg)
 \end{equation}
 \begin{equation}\label{eq:traj2}
\bm{v}(t) = \bigg(- \sigma r_{SG} \omega_c  \sin\omega_c t ,
  \sigma r_{SG} \omega_c ( -1 + \cos \omega_c t), 0\bigg)
 \end{equation}
which carve out cycloidal skipping orbits as shown in Figure~\ref{fig:geometry}.
The period is $T= 2 \pi /\omega_c$.
The periodicity of the skipping orbits is $\ell = -2 \pi r_{SG} \sigma$. 
From this solution, it is clear that opposite spins will separate from one another along the $y$-axis.
The average speed along the $y$-axis is $v_{avg} = \ell/T = -g \mu_B b\sigma/2 e B_0 = -r_{SG}\omega_c \sigma$ (and zero in $x$) which is identical to the quantum calculation.
This same average speed remains regardless of the initial position and velocity of the electrons. 
For an unpolarized electron spin system, the behavior is reminiscent of the spin Hall effect where a spin current is formed. However here, unlike the for the spin Hall effect, a longitudinal charge current and its concomitant dissipation is unnecessary \cite{Dyakonov2010}.

For non-ballistic transport, in the spirit of the Drude model we express the spin and charge dynamics in either the longitudinal or transverse geometry as
\begin{equation}
\frac{d\bm{p}}{dt} 
= -e \left(\bm{E}_{eff}\sigma  + \frac{\bm{p} \times \bm{B}_0}{m}\right) - \frac{\bm{p}}{\tau}
\end{equation}
with $E_{eff,i} = \frac{g\mu_B }{2e}\frac{\partial B_{\mathcal{Z},z}}{\partial r_i}$ being an effective electric field generated from a general Zeeman-only field. This effective field is constant and uniform though for the linear gradient assumed thus far which ensures analytic solutions.
A solution is readily found for each spin orientation in $z$, $\sigma$, in the steady state which gives for an unpolarized electron ensemble the second rank tensor of the spin current, $j_{i,z}$:
\begin{equation}\label{eq:spincurrentEQ}
j_{i,z} = \frac{g \mu_B}{2 e}\sigma_{ii}  \frac{\partial B_{\mathcal{Z},z}}{\partial r_i}\sigma  + \frac{g \mu_B}{2 e}\varepsilon_{izk} \sigma_{ik}\frac{\partial B_{\mathcal{Z},z}}{\partial r_k}\sigma
\end{equation}
with the conductivity tensor
\begin{equation}
 \hat{\bm{\sigma}}_c = \left( \begin{array}{ccc}
  \sigma_{xx} & -\sigma_{yx} & 0 \\
\sigma_{yx} & \sigma_{xx} &0 \\
0 & 0 & \sigma_{zz} \end{array} \right)
\end{equation}
where
\begin{equation}
\sigma_{0}  =  n e \mu , ~ \sigma_{xx} = \frac{\sigma_0}{1 + \omega_c^2 \tau^2},~ \sigma_{yx} = \sigma_{xx}\omega_c \tau, ~ \sigma_{zz} = \sigma_0
\end{equation}
and $\bm{B}_0 = B_0 \hat{z}$.
From this it is apparent that the charge current is zero, $j_c = j_+ + j_- = 0$ but the spin current,  $j_s = j_+ - j_- \neq 0$, is not.


\section{Discussion}
\begin{figure}[ptbh]
 \begin{centering}
        \includegraphics[scale = 0.5,trim = 280 365 0 90, angle = -0,clip]{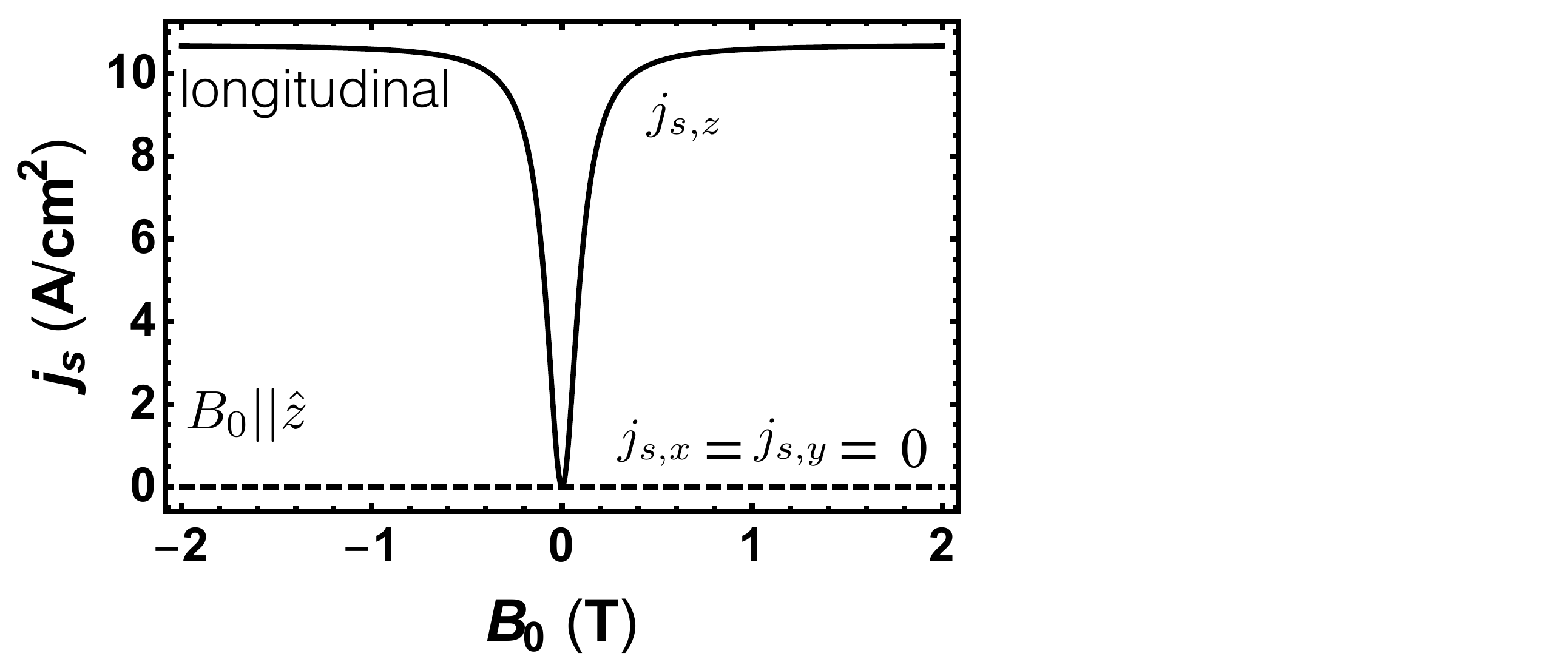}
        \caption[]
{Spin current density components versus applied magnetic field calculated in $n$-GaAs for a nuclear hyperfine gradient longitudinal ($\hat{z}$) to the applied field. Parameters: $\tau = 0.4$ ps, $n \approx 10^{16}$ cm$^{-3}$, $\sqrt{\xi}B_{\ell} = 100$ mT, $b_n = -1$ T, $\beta = 10^{-3}$ nm$^{-1}$ (corresponds to $b\approx -1$ mT/nm in a large field), and $g^* = -0.44$.}\label{fig:spincurrent2} 
        \end{centering}
\end{figure}

Now the nuclear field of Eq. (\ref{eq:zeemanfield}) is used for the Zeeman-only field and we make estimates of the spin current. 
In the longitudinal configuration, with $\hat{B}_0 || \hat{z}$, Eq. (\ref{eq:spincurrentEQ}) reduces to
\begin{equation}
\bm{j}_s  = 2 \sigma_{0} \bm{E}_{eff} = ne\mu \frac{g^* \mu_B}{e} \frac{ B_{0}^2}{B_{0}^2 + \xi B_{\ell}^2} b_n \beta \hat{B}_0 = n \mu g^* \mu_B b \hat{B}_0.
\end{equation}
This longitudinal spin current is plotted in Figure \ref{fig:spincurrent2}.
The width of the curve in Figure \ref{fig:spincurrent2} is governed by the local field, $B_{\ell}$.

\begin{figure}[ptbh]
 \begin{centering}
        \includegraphics[scale = 0.5,trim = 280 365 0 90, angle = -0,clip]{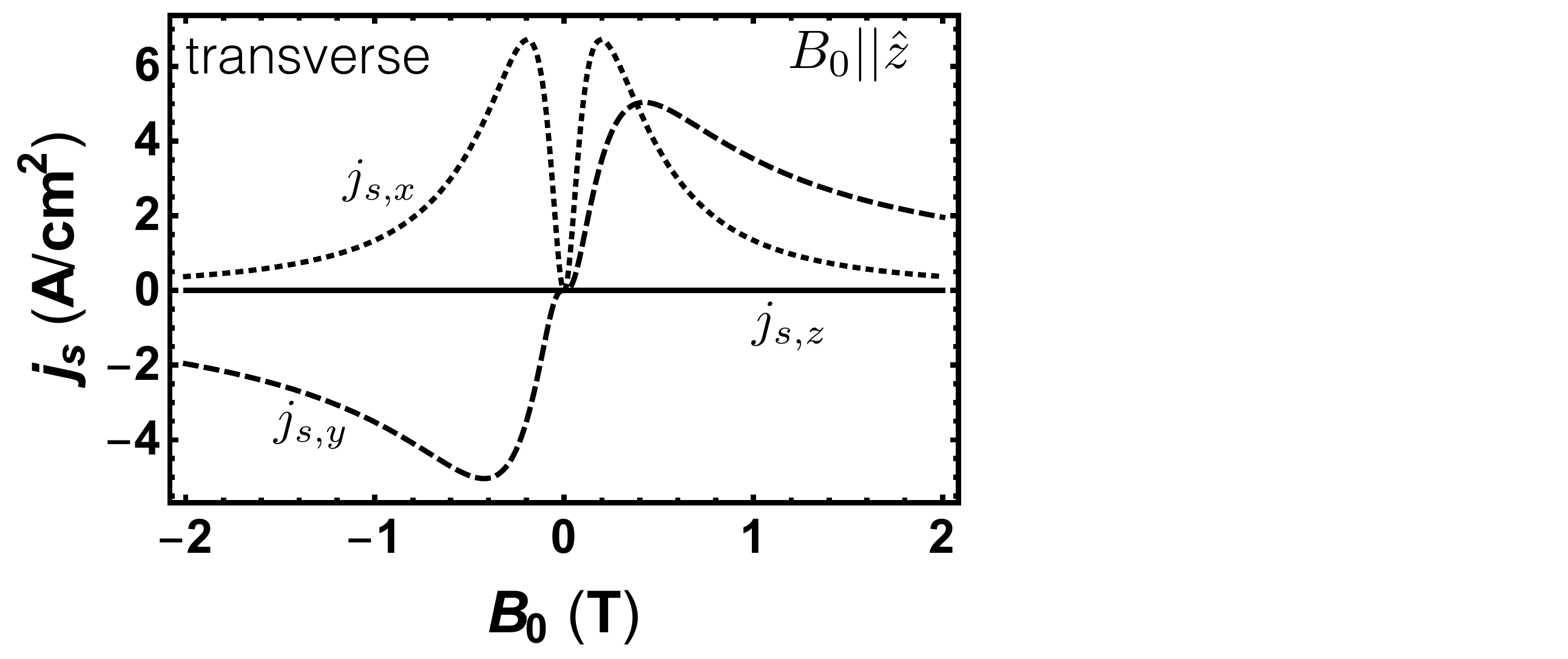}
        \caption[]
{Spin current density components versus applied magnetic field calculated in $n$-GaAs for a nuclear hyperfine gradient transverse ($\hat{x}$) to the applied field. Parameters: $\tau = 0.4$ ps, $n \approx 10^{16}$ cm$^{-3}$, $\sqrt{\xi}B_{\ell} = 100$ mT, $b_n = -1$ T, $\beta = 10^{-3}$ nm$^{-1}$ (corresponds to $b\approx -1$ mT/nm in a large field), and $g^* = -0.44$.}\label{fig:spincurrent} 
        \end{centering}
\end{figure}

Nuclear field gradients could be produced in a variety of ways. The simplest manner would be for the nuclear field to be graded by the inhomogeneous electron spin polarization arising from electron spin diffusion. In GaAs, the largest possible nuclear field is $\approx 17$ T which would correspond to efficient dynamic nuclear polarization from highly polarized electrons.\cite{Paget1977, Meier1984} In practice, the maximum nuclear field is smaller; Chan \emph{et al.} found it near 5 T.\cite{Chan2009}
At low temperatures, the spin diffusion length in doped GaAs is on the order of 10 $\mu$m. 
By ignoring additional nuclear spin diffusion, the decay of nuclear field follows that of the electron spin. If we take the maximum nuclear field slightly above 5 T, the $1/e$ field is about 2 T over 10 $\mu$m which leads to $b \approx 0.2$ mT/nm.
Further control of $b$ may be possible by controlling the electron spin diffusion length with an electric field.\cite{Yu2002, Yu2002a}

To find the size of longitudinal spin current to be expected in $n$-GaAs, we estimate the the conductivity to be $\sigma_0 = n e \mu = 2000/$$(\Omega$m)  with $n = 10^{16} ~$cm$^{-3}$ and $\mu = 10^4 ~$cm$^2$/Vs.
The effective `electric field' is determined by
\begin{equation}
E_{eff} = \frac{g^* \mu_B b}{2 e} = \frac{(-0.44)(9.3 \times 10^{-24}J/T)}{2 \times 1.6 \times 10^{-19}C} b 
\end{equation}
which computes to be $ 1.3 \times 10^{-5} b$ J/(T C)
where the $g$-factor for GaAs  $g^* =-0.44$.
Using $b = 0.2$ mT/nm, we find $E_{eff} \approx 3$ V/m. The spin current would be $2\sigma_0 E_{eff} = 1.2$ A/cm$^2$ which is comparable to values measured in the spin Hall mechanism.\cite{Garlid2010, Ehlert2012}

Larger linear effective field gradients may be possible by optically orienting \cite{Meier1984} spin with an appropriate optical grating. Through the process of dynamical nuclear polarization, an effective field is created parallel to the applied field with a transverse [Figure~\ref{fig:geometry}(a)] or longitudinal geometry.
The `slope' of the linear grating will dictate the strength of $\beta$ or $b$.
After generating the nuclear field and allowing the electronic spin to relax, an unpolarized pump can excite carriers that will undergo the dynamics described herein. 
A pure electron spin current will cross the sample. Note that the spin current is independent of the spin polarization. Kerr or Faraday rotation spectroscopy may then resolve the opposite spins on either side of the pump beam's spot. An alternate method of measurement, which may avoid charge recombination of spin carriers, is, after preparing the nuclear fields in the same manner, to have a polarized pump beam generate an imbalance of conduction electron spins which then result in a charge current, proportional to the injected spin polarization, to be measured at opposite contacts.

Our focus has been on the longitudinal spin current (as opposed to the transverse spin current) since it is able to achieve larger values in a broad field range.
For completeness, we  display the transverse spin current in Figure~\ref{fig:spincurrent}.
There are two field scales present: the narrow width is $\sim B_{\ell}$ and the larger width scales with the momentum relaxation rate.

 \section{Conclusion} 

In this article, we have demonstrated how nuclear fields, which are effective magnetic fields that do not affect orbital motion when uniform,  induce spin and charge currents when graded. Significant nuclear fields (on order of Tesla) are commonly created in doped GaAs which offers the chance to observe the effects described here. Managing gradients of these fields remains to be seen; we suggest an optical means by which dynamic nuclear polarization is filtered across a sample by selecting an appropriate optical grating. Our estimate of A/cm$^2$ spin current is similar to spin Hall currents measured in $n$-GaAs.

\section{Acknowledgements} This work was supported in part by the Center for Emergent Materials, an NSF MRSEC under Award No. DMR-1420451. NJH acknowledges additional support from the National Science Foundation under Grant Numbers DMR-2014786 and DMR-2152540.



\begin{thebibliography}{29}
\expandafter\ifx\csname natexlab\endcsname\relax\def\natexlab#1{#1}\fi
\expandafter\ifx\csname bibnamefont\endcsname\relax
  \def\bibnamefont#1{#1}\fi
\expandafter\ifx\csname bibfnamefont\endcsname\relax
  \def\bibfnamefont#1{#1}\fi
\expandafter\ifx\csname citenamefont\endcsname\relax
  \def\citenamefont#1{#1}\fi
\expandafter\ifx\csname url\endcsname\relax
  \def\url#1{\texttt{#1}}\fi
\expandafter\ifx\csname urlprefix\endcsname\relax\def\urlprefix{URL }\fi
\providecommand{\bibinfo}[2]{#2}
\providecommand{\eprint}[2][]{\url{#2}}

\bibitem[{\citenamefont{Awschalom et~al.}(2002)\citenamefont{Awschalom,
  Samarth, and Loss}}]{Awschalom2002}
\bibinfo{editor}{\bibfnamefont{D.~D.} \bibnamefont{Awschalom}},
  \bibinfo{editor}{\bibfnamefont{N.}~\bibnamefont{Samarth}}, \bibnamefont{and}
  \bibinfo{editor}{\bibfnamefont{D.}~\bibnamefont{Loss}}, eds.,
  \emph{\bibinfo{title}{Semiconductor Spintronics and Quantum Computation}}
  (\bibinfo{publisher}{Springer Verlag}, \bibinfo{address}{Heidelberg},
  \bibinfo{year}{2002}).

\bibitem[{\citenamefont{Awschalom and Flatt\'e}(2007)}]{Awschalom2007}
\bibinfo{author}{\bibfnamefont{D.~D.} \bibnamefont{Awschalom}}
  \bibnamefont{and} \bibinfo{author}{\bibfnamefont{M.~E.}
  \bibnamefont{Flatt\'e}}, \bibinfo{journal}{Nature Physics}
  \textbf{\bibinfo{volume}{3}}, \bibinfo{pages}{153} (\bibinfo{year}{2007}).

\bibitem[{\citenamefont{{D'Yakonov} and {Perel'}}(1971)}]{Dyakonov1971a}
\bibinfo{author}{\bibfnamefont{M.~I.} \bibnamefont{{D'Yakonov}}}
  \bibnamefont{and} \bibinfo{author}{\bibfnamefont{V.~I.}
  \bibnamefont{{Perel'}}}, \bibinfo{journal}{ZhETF Pisma Redaktsiiu}
  \textbf{\bibinfo{volume}{13}}, \bibinfo{pages}{657} (\bibinfo{year}{1971}).

\bibitem[{\citenamefont{D'yakonov and Perel'}(1971)}]{Dyakonov1971}
\bibinfo{author}{\bibfnamefont{M.~I.} \bibnamefont{D'yakonov}}
  \bibnamefont{and} \bibinfo{author}{\bibfnamefont{V.~I.}
  \bibnamefont{Perel'}}, \bibinfo{journal}{Physics Letters A}
  \textbf{\bibinfo{volume}{35}}, \bibinfo{pages}{459} (\bibinfo{year}{1971}).

\bibitem[{\citenamefont{Kato et~al.}(2004)\citenamefont{Kato, Myers, Gossard,
  and Awschalom}}]{Kato2004b}
\bibinfo{author}{\bibfnamefont{Y.~K.} \bibnamefont{Kato}},
  \bibinfo{author}{\bibfnamefont{R.~C.} \bibnamefont{Myers}},
  \bibinfo{author}{\bibfnamefont{A.~C.} \bibnamefont{Gossard}},
  \bibnamefont{and} \bibinfo{author}{\bibfnamefont{D.~D.}
  \bibnamefont{Awschalom}}, \bibinfo{journal}{Science}
  \textbf{\bibinfo{volume}{306}}, \bibinfo{pages}{1910} (\bibinfo{year}{2004}).

\bibitem[{\citenamefont{Wunderlich et~al.}(2005)\citenamefont{Wunderlich,
  Kaestner, Sinova, and Jungwirth}}]{Wunderlich2005}
\bibinfo{author}{\bibfnamefont{J.}~\bibnamefont{Wunderlich}},
  \bibinfo{author}{\bibfnamefont{B.}~\bibnamefont{Kaestner}},
  \bibinfo{author}{\bibfnamefont{J.}~\bibnamefont{Sinova}}, \bibnamefont{and}
  \bibinfo{author}{\bibfnamefont{T.}~\bibnamefont{Jungwirth}},
  \bibinfo{journal}{\prl} \textbf{\bibinfo{volume}{94}},
  \bibinfo{pages}{047204} (\bibinfo{year}{2005}).

\bibitem[{\citenamefont{Edelstein}(1990)}]{Edelstein1990}
\bibinfo{author}{\bibfnamefont{V.~M.} \bibnamefont{Edelstein}},
  \bibinfo{journal}{Solid State Communications} \textbf{\bibinfo{volume}{73}},
  \bibinfo{pages}{233} (\bibinfo{year}{1990}).

\bibitem[{\citenamefont{Ganichev et~al.}(2002)\citenamefont{Ganichev, Ivchenko,
  and Belkov}}]{Ganichev2002}
\bibinfo{author}{\bibfnamefont{S.}~\bibnamefont{Ganichev}},
  \bibinfo{author}{\bibfnamefont{E.}~\bibnamefont{Ivchenko}}, \bibnamefont{and}
  \bibinfo{author}{\bibfnamefont{V.}~\bibnamefont{Belkov}},
  \bibinfo{journal}{Nature} \textbf{\bibinfo{volume}{417}},
  \bibinfo{pages}{153} (\bibinfo{year}{2002}).

\bibitem[{\citenamefont{Rojas-Sanchez et~al.}(2013)\citenamefont{Rojas-Sanchez,
  Vila, and Desfonds}}]{Sanchez2013}
\bibinfo{author}{\bibfnamefont{J.-C.} \bibnamefont{Rojas-Sanchez}},
  \bibinfo{author}{\bibfnamefont{L.}~\bibnamefont{Vila}}, \bibnamefont{and}
  \bibinfo{author}{\bibfnamefont{G.}~\bibnamefont{Desfonds}},
  \bibinfo{journal}{Nat. Comm.} \textbf{\bibinfo{volume}{4}},
  \bibinfo{pages}{2944} (\bibinfo{year}{2013}).

\bibitem[{\citenamefont{Qi et~al.}(2006)\citenamefont{Qi, Yu, and
  Flatt\'e}}]{Qi2006a}
\bibinfo{author}{\bibfnamefont{Y.}~\bibnamefont{Qi}},
  \bibinfo{author}{\bibfnamefont{Z.-G.} \bibnamefont{Yu}}, \bibnamefont{and}
  \bibinfo{author}{\bibfnamefont{M.~E.} \bibnamefont{Flatt\'e}},
  \bibinfo{journal}{\prl} \textbf{\bibinfo{volume}{96}},
  \bibinfo{pages}{026602} (\bibinfo{year}{2006}).

\bibitem[{\citenamefont{Qi and Flatt\'e}(2019)}]{Qi2006b}
\bibinfo{author}{\bibfnamefont{Y.}~\bibnamefont{Qi}} \bibnamefont{and}
  \bibinfo{author}{\bibfnamefont{M.~E.} \bibnamefont{Flatt\'e}},
  \bibinfo{journal}{J. Supercon. Nov. Mag.} \textbf{\bibinfo{volume}{32}},
  \bibinfo{pages}{109} (\bibinfo{year}{2019}).

\bibitem[{\citenamefont{Petta et~al.}(2005)\citenamefont{Petta, Johnson,
  Taylor, Laird, Yacoby, Lukin, Marcus, Hanson, and Gossard}}]{Petta2005}
\bibinfo{author}{\bibfnamefont{J.~R.} \bibnamefont{Petta}},
  \bibinfo{author}{\bibfnamefont{A.~C.} \bibnamefont{Johnson}},
  \bibinfo{author}{\bibfnamefont{J.~M.} \bibnamefont{Taylor}},
  \bibinfo{author}{\bibfnamefont{E.~A.} \bibnamefont{Laird}},
  \bibinfo{author}{\bibfnamefont{A.}~\bibnamefont{Yacoby}},
  \bibinfo{author}{\bibfnamefont{M.~D.} \bibnamefont{Lukin}},
  \bibinfo{author}{\bibfnamefont{C.~M.} \bibnamefont{Marcus}},
  \bibinfo{author}{\bibfnamefont{M.~P.} \bibnamefont{Hanson}},
  \bibnamefont{and} \bibinfo{author}{\bibfnamefont{A.~C.}
  \bibnamefont{Gossard}}, \bibinfo{journal}{Science}
  \textbf{\bibinfo{volume}{309}}, \bibinfo{pages}{2180} (\bibinfo{year}{2005}).

\bibitem[{\citenamefont{Bobbert et~al.}(2007)\citenamefont{Bobbert, Nguyen, van
  Oost, Koopmans, and Wohlgenannt}}]{Bobbert2007}
\bibinfo{author}{\bibfnamefont{P.~A.} \bibnamefont{Bobbert}},
  \bibinfo{author}{\bibfnamefont{T.~D.} \bibnamefont{Nguyen}},
  \bibinfo{author}{\bibfnamefont{F.~W.~A.} \bibnamefont{van Oost}},
  \bibinfo{author}{\bibfnamefont{B.}~\bibnamefont{Koopmans}}, \bibnamefont{and}
  \bibinfo{author}{\bibfnamefont{M.}~\bibnamefont{Wohlgenannt}},
  \bibinfo{journal}{Phys. Rev. Lett.} \textbf{\bibinfo{volume}{99}},
  \bibinfo{pages}{216801} (\bibinfo{year}{2007}).

\bibitem[{\citenamefont{Harmon and Flatt\'e}(2012)}]{Harmon2012a}
\bibinfo{author}{\bibfnamefont{N.~J.} \bibnamefont{Harmon}} \bibnamefont{and}
  \bibinfo{author}{\bibfnamefont{M.~E.} \bibnamefont{Flatt\'e}},
  \bibinfo{journal}{Phys. Rev. Lett.} \textbf{\bibinfo{volume}{108}},
  \bibinfo{pages}{186602} (\bibinfo{year}{2012}).

\bibitem[{\citenamefont{Overhauser}(1953)}]{Overhauser1953}
\bibinfo{author}{\bibfnamefont{A.~W.} \bibnamefont{Overhauser}},
  \bibinfo{journal}{Phys. Rev.} \textbf{\bibinfo{volume}{92}},
  \bibinfo{pages}{411} (\bibinfo{year}{1953}).

\bibitem[{\citenamefont{Meier and Zachachrenya}(1984)}]{Meier1984}
\bibinfo{author}{\bibfnamefont{F.}~\bibnamefont{Meier}} \bibnamefont{and}
  \bibinfo{author}{\bibfnamefont{B.~P.} \bibnamefont{Zachachrenya}},
  \emph{\bibinfo{title}{Optical Orientation: Modern Problems in Condensed
  Matter Science}}, vol.~\bibinfo{volume}{8}
  (\bibinfo{publisher}{North-Holland}, \bibinfo{address}{Amsterdam},
  \bibinfo{year}{1984}).

\bibitem[{\citenamefont{Bednarski}(1998)}]{Bednarski1998}
\bibinfo{author}{\bibfnamefont{H.}~\bibnamefont{Bednarski}},
  \bibinfo{journal}{Physica B} \textbf{\bibinfo{volume}{258}},
  \bibinfo{pages}{641} (\bibinfo{year}{1998}).

\bibitem[{\citenamefont{Vagner}(2004)}]{Vagner2004}
\bibinfo{author}{\bibfnamefont{I.~D.} \bibnamefont{Vagner}},
  \bibinfo{journal}{HAIT Journal of Science and Engineering}
  \textbf{\bibinfo{volume}{1}}, \bibinfo{pages}{152} (\bibinfo{year}{2004}).

\bibitem[{\citenamefont{Harmon et~al.}(2015)\citenamefont{Harmon, Peterson,
  Geppert, Patel, Palmstr{\o}m, Crowell, and Flatt{\'{e}}}}]{Harmon2015}
\bibinfo{author}{\bibfnamefont{N.~J.} \bibnamefont{Harmon}},
  \bibinfo{author}{\bibfnamefont{T.~A.} \bibnamefont{Peterson}},
  \bibinfo{author}{\bibfnamefont{C.~C.} \bibnamefont{Geppert}},
  \bibinfo{author}{\bibfnamefont{S.~J.} \bibnamefont{Patel}},
  \bibinfo{author}{\bibfnamefont{C.~J.} \bibnamefont{Palmstr{\o}m}},
  \bibinfo{author}{\bibfnamefont{P.~A.} \bibnamefont{Crowell}},
  \bibnamefont{and} \bibinfo{author}{\bibfnamefont{M.~E.}
  \bibnamefont{Flatt{\'{e}}}}, \bibinfo{journal}{Physical Review B}
  \textbf{\bibinfo{volume}{92}}, \bibinfo{pages}{140201}
  (\bibinfo{year}{2015}).

\bibitem[{\citenamefont{Ou et~al.}(2016)\citenamefont{Ou, Chiu, Harmon,
  Odenthal, Sheffield, Chilcote, Kawakami, and Flatt{\'{e}}}}]{Ou2016}
\bibinfo{author}{\bibfnamefont{Y.-S.} \bibnamefont{Ou}},
  \bibinfo{author}{\bibfnamefont{Y.-h.} \bibnamefont{Chiu}},
  \bibinfo{author}{\bibfnamefont{N.~J.} \bibnamefont{Harmon}},
  \bibinfo{author}{\bibfnamefont{P.}~\bibnamefont{Odenthal}},
  \bibinfo{author}{\bibfnamefont{M.}~\bibnamefont{Sheffield}},
  \bibinfo{author}{\bibfnamefont{M.}~\bibnamefont{Chilcote}},
  \bibinfo{author}{\bibfnamefont{R.~K.} \bibnamefont{Kawakami}},
  \bibnamefont{and} \bibinfo{author}{\bibfnamefont{M.~E.}
  \bibnamefont{Flatt{\'{e}}}}, \bibinfo{journal}{Phys. Rev. Lett.}
  \textbf{\bibinfo{volume}{116}}, \bibinfo{pages}{107201}
  (\bibinfo{year}{2016}).

\bibitem[{\citenamefont{Paget et~al.}(1977)\citenamefont{Paget, Lampel,
  Sapoval, and Safarov}}]{Paget1977}
\bibinfo{author}{\bibfnamefont{D.}~\bibnamefont{Paget}},
  \bibinfo{author}{\bibfnamefont{G.}~\bibnamefont{Lampel}},
  \bibinfo{author}{\bibfnamefont{B.}~\bibnamefont{Sapoval}}, \bibnamefont{and}
  \bibinfo{author}{\bibfnamefont{V.~I.} \bibnamefont{Safarov}},
  \bibinfo{journal}{Phys. Rev. B} \textbf{\bibinfo{volume}{15}},
  \bibinfo{pages}{5780} (\bibinfo{year}{1977}).

\bibitem[{\citenamefont{Muller}(1992)}]{Muller1992}
\bibinfo{author}{\bibfnamefont{J.~E.} \bibnamefont{Muller}},
  \bibinfo{journal}{Phys. Rev. Lett.} \textbf{\bibinfo{volume}{68}},
  \bibinfo{pages}{385} (\bibinfo{year}{1992}).

\bibitem[{\citenamefont{Berggren et~al.}(1988)\citenamefont{Berggren, Roos, and
  van Houten}}]{Berggren1988}
\bibinfo{author}{\bibfnamefont{K.~F.} \bibnamefont{Berggren}},
  \bibinfo{author}{\bibfnamefont{G.}~\bibnamefont{Roos}}, \bibnamefont{and}
  \bibinfo{author}{\bibfnamefont{H.}~\bibnamefont{van Houten}},
  \bibinfo{journal}{Phys. Rev. B} \textbf{\bibinfo{volume}{37}},
  \bibinfo{pages}{118} (\bibinfo{year}{1988}).

\bibitem[{\citenamefont{Dyakonov}(2010)}]{Dyakonov2010}
\bibinfo{author}{\bibfnamefont{M.~I.} \bibnamefont{Dyakonov}}, in
  \emph{\bibinfo{booktitle}{Future Trends in Microelectronics: From
  Nanophotonics to Sensors to Energy}}, edited by
  \bibinfo{editor}{\bibfnamefont{S.}~\bibnamefont{Luryi}},
  \bibinfo{editor}{\bibfnamefont{J.}~\bibnamefont{Xu}}, \bibnamefont{and}
  \bibinfo{editor}{\bibfnamefont{A.}~\bibnamefont{Zaslavsky}}
  (\bibinfo{publisher}{John Wiley and Sons, Hoboken, New Jersey},
  \bibinfo{year}{2010}), p. \bibinfo{pages}{251}.

\bibitem[{\citenamefont{Chan et~al.}(2009)\citenamefont{Chan, Hu, Zhang, Kondo,
  Palmstr{\o}m, and Crowell}}]{Chan2009}
\bibinfo{author}{\bibfnamefont{M.~K.} \bibnamefont{Chan}},
  \bibinfo{author}{\bibfnamefont{Q.~O.} \bibnamefont{Hu}},
  \bibinfo{author}{\bibfnamefont{J.}~\bibnamefont{Zhang}},
  \bibinfo{author}{\bibfnamefont{T.}~\bibnamefont{Kondo}},
  \bibinfo{author}{\bibfnamefont{C.~J.} \bibnamefont{Palmstr{\o}m}},
  \bibnamefont{and} \bibinfo{author}{\bibfnamefont{P.~A.}
  \bibnamefont{Crowell}}, \bibinfo{journal}{Phys. Rev. B}
  \textbf{\bibinfo{volume}{80}}, \bibinfo{pages}{161206(R)}
  (\bibinfo{year}{2009}).

\bibitem[{\citenamefont{Yu and Flatt{\'{e}}}(2002)}]{Yu2002}
\bibinfo{author}{\bibfnamefont{Z.~G.} \bibnamefont{Yu}} \bibnamefont{and}
  \bibinfo{author}{\bibfnamefont{M.~E.} \bibnamefont{Flatt{\'{e}}}},
  \bibinfo{journal}{Physical Review B} \textbf{\bibinfo{volume}{66}},
  \bibinfo{pages}{235302} (\bibinfo{year}{2002}), \eprint{0206321}.

\bibitem[{\citenamefont{Yu and Flatt\'e}(2002)}]{Yu2002a}
\bibinfo{author}{\bibfnamefont{Z.~G.} \bibnamefont{Yu}} \bibnamefont{and}
  \bibinfo{author}{\bibfnamefont{M.~E.} \bibnamefont{Flatt\'e}},
  \bibinfo{journal}{Phys. Rev. B} \textbf{\bibinfo{volume}{66}},
  \bibinfo{pages}{201202} (\bibinfo{year}{2002}).

\bibitem[{\citenamefont{Garlid et~al.}(2010)\citenamefont{Garlid, Hu, Chan,
  Palmstr{\o}m, and Crowell}}]{Garlid2010}
\bibinfo{author}{\bibfnamefont{E.~S.} \bibnamefont{Garlid}},
  \bibinfo{author}{\bibfnamefont{Q.~O.} \bibnamefont{Hu}},
  \bibinfo{author}{\bibfnamefont{M.~K.} \bibnamefont{Chan}},
  \bibinfo{author}{\bibfnamefont{C.~J.} \bibnamefont{Palmstr{\o}m}},
  \bibnamefont{and} \bibinfo{author}{\bibfnamefont{P.~A.}
  \bibnamefont{Crowell}}, \bibinfo{journal}{Physical Review Letters}
  \textbf{\bibinfo{volume}{105}}, \bibinfo{pages}{156602}
  (\bibinfo{year}{2010}).

\bibitem[{\citenamefont{Ehlert et~al.}(2012)\citenamefont{Ehlert, Song, Ciorga,
  Utz, Schuh, Bougeard, and Weiss}}]{Ehlert2012}
\bibinfo{author}{\bibfnamefont{M.}~\bibnamefont{Ehlert}},
  \bibinfo{author}{\bibfnamefont{C.}~\bibnamefont{Song}},
  \bibinfo{author}{\bibfnamefont{M.}~\bibnamefont{Ciorga}},
  \bibinfo{author}{\bibfnamefont{M.}~\bibnamefont{Utz}},
  \bibinfo{author}{\bibfnamefont{D.}~\bibnamefont{Schuh}},
  \bibinfo{author}{\bibfnamefont{D.}~\bibnamefont{Bougeard}}, \bibnamefont{and}
  \bibinfo{author}{\bibfnamefont{D.}~\bibnamefont{Weiss}},
  \bibinfo{journal}{Physical Review B} \textbf{\bibinfo{volume}{86}},
  \bibinfo{pages}{205204} (\bibinfo{year}{2012}).

\end{thebibliography}









\end{document}